*Chapter 4*

# Managing health insurance using blockchain technology

*Tajkia Nuri Ananna[1], Munshi Saifuzzaman[2], Mohammad Jabed Morshed Chowdhury[3] and Md Sadek Ferdous[4]*

## Abstract

Health insurance plays a significant role in ensuring quality healthcare. In response to the escalating costs of the medical industry, the demand for health insurance is soaring. Additionally, those with health insurance are more likely to receive preventative care than those without health insurance. However, from granting health insurance to delivering services to insured individuals, the health insurance industry faces numerous obstacles. Fraudulent actions, false claims, a lack of transparency and data privacy, reliance on human effort and dishonesty from consumers, healthcare professionals, or even the insurer party itself, are the most common and important hurdles towards success. Given these constraints, this chapter briefly covers the most immediate concerns in the health insurance industry and provides insight into how blockchain technology integration can contribute to resolving these issues. This chapter finishes by highlighting existing limitations as well as potential future directions.

**Keywords:** Blockchain; Health insurance management; Customer service management; Fraud detection and risk prevention; Claim and billing management; Data record and sharing management

## 4.1 Introduction

Health insurance is a means of providing financial support for a person's healthcare expenses [1]. In exchange for a premium, the insured must pay all medical, surgical, and even dental treatment charges incurred by the insurer. Many organizations provide health insurance benefits to their employees, which encourages them to perform well [2]. Accidents and illnesses are unpredictable in life because they can happen to anyone at any time. The unexpected cost of treatment can put a huge strain on a person's finances. Healthcare costs are increasing rapidly as medical technology advances [3,4]. Because medical costs are going up and people want better care, more people are getting health insurance [5]. According to the study by Mendoza-Tello *et al.* [5], health insurance acts as a shield of safety in three ways. To begin with, it ensures preventative access to medicine and healthcare. When compared to uninsured individuals, those with health insurance have a better chance of accessing early detection tests for any illness. In the United States of America, lack of health insurance causes poor cancer outcomes. Uninsured individuals do not receive preventative care, which results in delayed cancer detection and a lower chance of survival [6–9]. Uninsured people are also more likely to be diagnosed at a later stage of the disease [10,11] and have a lower chance of surviving [8,11]. Second, it enables access to disease treatment that is personalized to the insured's preferences by the insurer. Insured individuals are more likely to obtain evidence-based care than uninsured individuals [9,12]. Third, it gives economic protection, discounts, and compensation for health care expenses. Insurance companies negotiate discounts with healthcare providers and thus pay a major portion of the insured's medical expenses [13]. That is how an insured person has access to reimbursement and financial stability compared to an uninsured person. An uninsured person may be forced to pay a large sum of money, resulting in a significant financial loss and health deterioration [5,14]. The health insurance issuance process starts when the customer fills out the application form in order to purchase a plan [5,15]. The insurance company that provides insurance service is addressed as the insurer. A person who purchases insurance is called an insured party. Depending on the insured's selected plan, medical background, age and insurance sum, the insurer offers the premium amount, which is paid by the insured party to the insurer [5,15]. In some situations, the insured party is required to undergo a medical examination before the insurer decides how much money to give the insured party [15]. Finally, the insurer is provided a customized policy and receives health coverage based on a mutual contract between the insurer and healthcare provider [5,15].

[1]Department of Computer Science and Engineering, Metropolitan University, Sylhet, Bangladesh
[2]Dynamic Solution Innovators Ltd, Dhaka, Bangladesh
[3]Department of Computer Science and IT, La Trobe University, Australia
[4]Department of Computer Science and Engineering, BRAC University, Bangladesh

The health insurance industry's processes are riddled with inefficiencies, such as claim falsification, duplicate claims, fraud activities, irregular billing, security and privacy difficulties in data sharing and so on. There are numerous areas that require development in terms of efficiency, accuracy, and management. Fraud in health insurance is one of the major concerns among the issues. On September 30, 2020, healthcare insurance fraud to federal healthcare programs and private insurers have resulted in a $6 billion loss in the United States. A total of 345 individuals have charged for their roles in this fraud, with 100 of them being doctors, nurses, and medical professionals [16]. According to a survey by the National Health Care Anti-Fraud Association (NHCAA), one healthcare costs the United States more than $2.27 trillion each year, with tens of billions of dollars lost due to healthcare insurance fraud [17]. Security and privacy issues have been a major concern when it comes to health data sharing as people are extremely sensitive about their medical data, which makes customers hesitant to disclose their health data. Many people believe that the healthcare industry is not well prepared to deal with the growing number of cyber threats [18].

Various fields, such as machine learning, blockchain, and others, have had an effect on health insurance management. Machine learning is one of the most important and rapidly developing fields in the world today. Although machine learning has not been utilized extensively in the management of health insurance, there have been a few noteworthy achievements in this area. The authors of [19] have proposed an approach based on machine learning to combat health insurance fraud. Similarly, machine learning implementations have been observed in [20] and [21]. In [22], authors have leveraged data mining technologies to address fraud and claim management problems in health insurance management whereas authors of [23] have made notable contributions to the resolution of claim processing in health management. However, the most significant issue with all of these works is that privacy and security have not been adequately addressed. In addition, no mechanism for the secure exchange of data has been presented. This is a vulnerable case because privacy is the top priority for the vast majority of people today. Blockchain, on the other hand, is one of the most secure emerging technologies, and it possesses a number of characteristics that allow it to solve these significant issues. Blockchain is a distributed, decentralized, and immutable ledger technology that facilitates the transaction recording process and assists in the tracking of both tangible (cash, land) and intangible assets (intellectual property, patents, copyrights) [24,25]. It acts as a database, storing information electronically in a digital format [26]. Each transaction in a blockchain is recorded as blocks that are linked together using cryptography [27–29]. The connection between blocks provides the blockchain's immutability or resistance to manipulation by assuring that no blocks can be altered [24]. It provides immediate, shared, and completely transparent information that is stored on an immutable ledger that is accessible only to permissioned network members [24]. Blockchain technology has the characteristics of decentralization, tamper-resistance, and traceability [30]. It is capable of constructing a secure and private network as well as has the potential to resolve a wide range of health insurance issues such as interoperability, fraud activities, billing, service management and so on. Additionally, smart contracts on the blockchain can make the entire insurance process and documentation transparent. Blockchain holds all parties, including healthcare professionals, pharmaceutical companies, and insurance companies, accountable for their activities. This contributes to the development of trust among insurers and policyholders, healthcare professionals and patients as well as the overall healthcare industry [31]. As a result, it can be stated that blockchain has the capability of resolving the majority of the concerns inherent in the current health insurance system and creating a secure environment to facilitate the entire health insurance system.

This chapter reviews the most promising research work on the management of health insurance using blockchain technology. To the best of knowledge, there is not a single review paper, survey study, or even book chapter that can assist future scholars in this topic, which motivates this work. In order to classify existing works, this chapter divides health insurance management into a number of key research areas. Consequently, this would assist future researchers get a broad understanding of the field and figure out where blockchain and the existing literature fall short.

**Authors' contributions:** The authors' contributions in this chapter are as follows:

1. A survey on managing health insurance using blockchain technology has been conducted. This chapter assesses the techniques, their use cases, experiment scenarios, and their limitations by looking at the most relevant and recent research studies.
2. Five different aspects are then analyzed by presenting foundational demonstrations and major key concepts from existing works, and a summary was drawn.
3. This chapter concludes with a comprehensive discussion of the limitations and future directions of the reviewed papers.

The following is how this chapter is organized: it begins with Section 4.2, which discusses several key aspects and challenges of the health insurance industry. The causal effects of health insurance then are discussed in Section 4.3. Section 4.4 discusses blockchain's revolutionary impact on the health insurance industry. There is a detailed discussion of a few of the most important core factors that affect health insurance management from Section 4.5 to Section 4.9. The chapter concludes with a discussion of limitations and potential future scopes focused on the reviewed literature as well as blockchain technology in Section 4.10.

## 4.2 The insurance industry: its aspects and challenges

Insurance, in its simplest form, protects any individual, organization, or other entity (entity in short) against financial loss. The purpose of insurance is to safeguard any entity's financial well-being in the event of an unforeseen misfortune for instance, structural damage, loss of well-being, and so forth [32].

The insurance industry is vast, having numerous components intertwined with it. It is one of the areas where customer service management plays an important role in the overall process. If the consumer is dissatisfied with the service, the entire system suffers substantial losses. The insurance company should be able to handle the insured parties' complaints and grievances, as they play a significant role in the survival of any insurance company. Given all of these considerations, insurance should be an honest field with the goal of selflessly assisting people. But, in reality, the situation is considerably different from what was anticipated. The conviction about insurance is that if a person receives insurance and pays the monthly premium, he or she will not have to worry about paying in the event of an unexpected accident or distress. That is the insurance system's expected behavior. However, despite all of its benefits, the insurance industry has so many dark and sophisticated sides that it is difficult to comprehend. There are so many areas where the insurance sector gets compromised by being falsely used by fraud activities. The actors behind these activities can be the customer, the healthcare provider or even the insurer themselves. There are many ways the insurance sector can be compromised by an insured person or healthcare provider, which are as follows:

1. The insured party falsely claiming for additional financial assistance from the insurance agency.
2. Insured individual claiming money from the insurer for services that were never provided.
3. Falsifying facts while issuing insurance cards.
4. Healthcare providers may charge for irrelevant or unprovided services.
5. Extra money can be claimed by drugstores and healthcare providers for medicine and other items.

Insurance agencies can do many things to avoid responsibility and overcharge the customer with higher rates to tactfully gain more cash. The most common medical coverage scams that can deceive customers are listed below [33]:

1. Insurance companies may purposefully deny claims in order to create a scenario in which they are the victim and the healthcare professionals are the scammers.
2. The insurance company can change the coverage without notice in order to further their selfish desires.
3. They may charge the insured for out-of-network costs that the insured is unaware of, or they may add hidden costs to the insurance premium.

Along with these, there is widespread concern among consumers regarding the security and privacy of health data shared with insurance companies, as people are more concerned about their privacy. The method of exchanging information is inefficient and does not adequately protect individuals' privacy. Additionally, the current insurance system is so reliant on manual work that it exposes the entire system to damage. As a result, all critical aspects of health insurance require immediate attention. The core aspects of the health insurance industry are illustrated in Figure 4.1.

## 4.3 Causality in health insurance

In general, "casualty" refers to the link between the cause and the impact of a specific event or occurrence. Consequently, it is extremely significant in terms of health insurance. The effect of the presence or absence of health insurance on healthcare utilization and health outcomes is referred to as a casualty in health insurance. The most important elements influencing healthcare advancement are optimized access to healthcare and increased quality of care delivered [34–37]. One of the most concerning topics in the United States has been the lack of health insurance and its potential impact on healthcare utilization and health outcomes [38]. The New York Times reported a list of nearly 47 million uninsured people in the United States in 2006 [39]. In addition, the Institute of Medicine [40] and the National Coalition on Health Care [41] have recommended universal health coverage to prevent a potential health disaster.

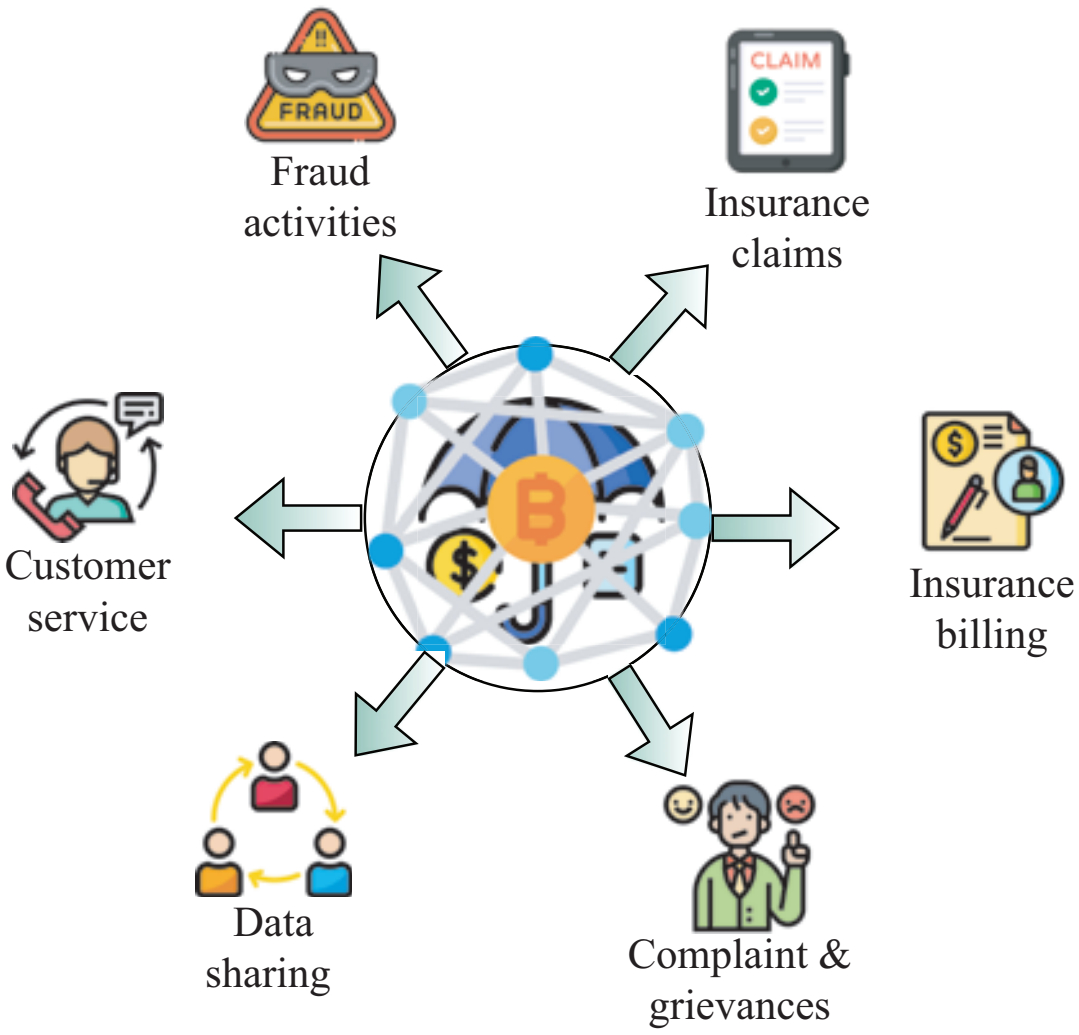

*Figure 4.1 Health insurance aspects*

Though most people, including citizens and politicians, believe that universal health insurance coverage is extremely important, relatively few actions have been taken to bring about this basic and significant change in practice. Few studies have addressed the relation between health insurance and healthcare access, utilization and health outcome [38].

The American college of Physicians – American society of internal medicine [42] presents the significant data it has collected in "No health insurance? It is enough to make you sick." According to the findings, insured people have more access to healthcare, receive better hospital-based care, have a lower likelihood of having delayed treatment, and have a lower overall death rate. Brown *et al.* [43] have also summarized their work in the same way, demonstrating that insured people have better access to healthcare, resulting in a reduced mortality rate. Hadley *et al.* [44] likewise have emphasized that insured people have better access to healthcare, preventative and diagnostic services, and have lower mortality rates. Another key aspect has been discovered: insured people are more likely to be diagnosed with any illness at an early stage. Hoffman and Paradise [45] have highlighted the relationship between health insurance and health outcomes. It has been observed that insured people receive better care for chronic diseases and are less likely to encounter the obstacles of delayed treatment or missing out on necessary treatment. Howell *et al.* [46] have demonstrated the benefits of health insurance coverage for pregnant women in terms of better health outcomes before, during, and after the baby is born. Insured people also have the chance of getting regular checkups from a doctor and have a lower waiting time [49]. Table 4.1 illustrates some of the studies that have focused on analyzing the relationship between health insurance, healthcare utilization, and health outcomes.

Furthermore, it is obvious from the preceding discussion that there is a link between health insurance and healthcare utilization and/or health outcomes. When compared to uninsured people, insured people receive better treatment and facilities, as well as more preventative care. All of these amenities contribute to lower mortality rates since insured people receive preventative treatment, have better

*Table 4.1 Research findings on the effects of health insurance*

| **References** | **Summary**[a] |
|---|---|
| American College of Physicians: no health insurance? It is enough to make you sick [42] | Insured are more likely to have proper access to healthcare and preventative services, they are less likely to receive delayed treatment, report mismanagement or missing services, receive better hospital-based care, and have a lower mortality rate |
| Brown *et al.* [43] | Insured have better access to healthcare and lower mortality rate |
| Hadley [44] | Insured receive better preventative service and access to medical care, more likely to be diagnosed at an early stage of any disease and have lower mortality rates |
| Hoffman and Paradise [45] | Insured have greater access to care, less likely to encounter any unmet treatment or delayed treatment, access better hospitalized care and lower mortality rate |
| Howell [46] | Increased health coverage for pregnant women before, during and after birth |
| Institute of Medicine: care without coverage [47] | Insurance provides regular healthcare facilities, an increase in preventative treatments, and proper treatment for chronic illnesses, as well as improved health outcomes |
| Office of Technology Assessment: does health insurance make a difference? [48] | Insured people are more likely to receive proper care and have fewer negative health outcomes |
| Weissman and Epstein [49] | Insured people are more likely to have a regular doctor and to obtain preventive care. They have shorter treatment wait times, are less likely to receive delayed care, and obtain better emergency care |

[a]Conclusions are presented in comparison to the uninsured, with the exception of Howell (2001) and Levy and Meltzer (2001)

facilities, and can be detected at an earlier stage of any disease. As a result, health insurance improves health outcomes and healthcare utilization.

## 4.4 Blockchain technology: a revolution in the health insurance industry

As discussed previously, the health insurance industry is being compelled to compromise its efficiency and prosperity as a result of the high number of hazardous incidents. Blockchain technology possesses the potential to bring numerous benefits and resolve the majority of health insurance problems and concerns. The health insurance industry is being disrupted by blockchain innovation and this time for good! According to Markets and Markets [50], the global market for blockchain in insurance is predicted to grow from USD 64.50 million in 2016 to USD 1,393.8 million in 2025 [51].

So, on that basis, here are a few important benefits of blockchain in the health insurance industry:

- **Efficient data sharing:** One of the most crucial components of the blockchain is its transparency. The blockchain provides an immutable and transparent data exchange technology that assures the integrity of the data shared through the blocks. It aids insurance companies in ensuring the authenticity of shared data. Blockchain is protected by cryptographic techniques such as digital signatures and hash functions, which safeguard the process of data sharing and protect the privacy of each individual.
- **Combating frauds and false claims:** One of the most concerning hazards in the health insurance industry is false claims and fraudulent activities. The health insurance industry takes so many preventative measures in order to avoid false claims and fraudulent activities. However, the adversaries manage to deceive the insurance industry in some way. As a result of the Blockchain's ability to record time-stamped transactions with complete audit trials, counterfeiters find it incredibly hard to commit fraud [51].
- **Enhancement in customer experience:** Customer satisfaction is a critical component of any insurance company's sustainability. Customers are typically content with the provider for whom they pay lower premiums and receive enhanced benefits. Additionally, if an insurance agency is capable of responding quickly to customer concerns and providing justice to any client, this contributes significantly to strengthening consumer trust in that insurance company. However, maintaining all of these simultaneously is extremely difficult. One solution is to leverage blockchain technology to automate processing through the use of smart contracts. Business agreements are encoded in the blockchain, and payments are triggered and processed instantly [51].
- **Improves trust among entities:** In this scenario, the blockchain smart contract emerges as a savior by establishing trust throughout the formation of the insurance contract. In every transaction, smart contracts provide immutability and auditability. In addition, it eliminates the need for an intermediary to manage the insurance process.

- **Collect and store data:** The insurance industry is data-driven. Using technologies such as artificial intelligence (AI) and the Internet of Things (IoT), blockchain technology enables the collection of a wide variety of valuable data. IoT-collected data is stored on the blockchain and then analyzed by AI. This enables the company to take an informed and autonomous decision on insurance premiums [51,52].
- **Accountability and ownership:** Accountability allows all parties involved in a commitment to be open and honest with one another. The immutability of block transactions and their connections enables ownership control and accountability [53]. Using the open and decentralized nature of blockchain, customers can see what the insurance company knows about them and how that information is used, which means that there is more openness and honesty [54].

According to Kuo *et al*. [55], the major benefit of blockchain integration for the healthcare industry is the decentralization behavior of blockchain, which allows relevant parties such as healthcare providers, insurance companies, and government regulatory bodies to share data securely without relying on any central system. Immutability, transparency, security, and robustness are other important factors in making any blockchain-based system more efficient and improving data sharing. With such exceptional benefits, it may be concluded that blockchain is the optimal way to help the health insurance industry. In later sections, the most crucial aspects of health insurance are investigated thoroughly, along with how blockchain can provide a substantial path to resolving these issues.

## 4.5 Customer service management of health insurance using blockchain technology

Customer service is critical in developing a positive relationship between a company and its customers. To be specific, a company's success is heavily reliant on customer service and satisfaction. Customer service is the assistance provided by enterprises before, during, and after the purchase or use of their products or services. Medical aid supplied to patients, any type of insurance service provided by insurance firms, or any materialistic product sold by companies are all instances of customer service. Good customer service improves customer satisfaction and enterprise performance, which leads to the enterprise's ultimate success [56].

The quality of customer service is not just determined by the customer service. It is heavily dependent on how the organization uses its current customer service scenario to attract new customers and retain the loyalty of existing ones [57]. Customer service is a kind of knowledge-based work where the main motive is to efficiently utilize and accumulate previous knowledge to improve the customer service quality [58]. Therefore, it can be stated that an efficient customer service management system is an essential component of every company's success.

Health insurance provides customer service to its members who purchase health insurance from their company. The objective of a health insurance company is to provide financial benefits to its customers. An insured person receives health coverage as a result of the insurer's and health-care providers' combined efforts. However, managing customers and providing adequate customer care is a difficult task. The entire system is entirely dependent on human supervision, making it prone to error. Other drawbacks include the insured individual fabricating information, abnormal billing, a lack of transparency, and, most crucially, a lack of data privacy and security. These difficulties must be addressed in order for the health insurance organization to be successful. Blockchain has the ability to alleviate the majority of the problems associated with the health insurance customer administration system. Blockchain technology can provide a secure environment in which the health insurance management system can be easily integrated.

According to the preceding discussion, customer service management is about providing quality service to existing customers as well as attracting new customers with better service. Considering this information, it can be stated that there are some factors that influence quality customer management, such as thoroughly monitoring if current customers are receiving proper facilities; how the insurance company responds in the event of any obstacles encountered by the customer; and, most importantly, if any customer's complaint is being resolved patiently. As a result, customer management in health insurance falls into two broad categories: *monitor and response management* and *complaint and grievance Management*. These two aspects are among the core aspects of health insurance management, as demonstrated in the following sub sections.

### *4.5.1 Monitor and response management*

Monitor and response management in health insurance play a vital role in ensuring quality customer service. Monitor and response management in health insurance refers to the ability to properly manage health insurance through monitoring and ensure better quality service by providing a proper response to customer demands and issues.

The insurance industry's major step in this regard is to constantly monitor all of their customer facilities, such as whether customers are receiving all of the services included in their insurance policy. If this factor is not ensured, achieving existing customer satisfaction can be extremely difficult, let alone attracting new customers. However, this aspect is more than just constantly monitoring existing client facilities; it serves a larger purpose. With medical fields getting better and new cures and devices being made almost every day, the cost of medical care and getting access to it has become hard to handle. As a result, insurance companies must monitor their policies on a regular basis and, if necessary, reform them. Otherwise, maintaining the standard and remaining among the best companies is extremely difficult.

The main challenge in ensuring these aforementioned factors is ensuring quality service by providing adequate privacy to customers. As monitoring and response management are essentially about sharing information about customers, it is critical to protect each customer's privacy. Customers must also be assured that their privacy is not violated. Current systems, however, struggle to meet these requirements due to inconsistencies in security policies and access control structures. Blockchain paves the way for a revolution in this field. If the privacy and transparency properties of blockchain can be efficiently integrated into solving these issues, it could be a massive benefit and make the monitoring and response management process run smoothly.

### *4.5.2 Complaint and grievances management*

It is crucial how a health insurer handles customer complaints and concerns. The quality of their service is highly dependent on how they handle their customers' concerns [59]. Customer complaint resolution facilitates the resolution of unsatisfactory situations and the tracking of complaints and grievances, paving the way for an organization to enhance its service quality [60,61].

Any individual has the freedom to select any insurer he or she wishes, based on the insurer's service quality. Customers have the option of switching plans or even insurance companies if they are displeased with the outcome of their complaints [62]. In typical health insurance complaint and grievance management systems, the client must physically attend in order to file a complaint or even monitor its status. Even if the customer's complaint is addressed, he or she might be dissatisfied with the entire system if the settlement is too lengthy. Additionally, there is the concern of transparency and anonymity. In today's society, people are extremely protective of their data, which makes them feel uncomfortable filing a complaint just because the system lacks anonymity. There may be additional causes for ineffective complaint and grievance management, such as a failure to enforce robust complaint and grievance management laws and regulations or a flawed redressal system [63].

To the best of our knowledge, there is no literature that directly addresses health insurance complaints and grievances. As a result, similar fields have been investigated how research communities have addressed complaints and grievances. Health insurance companies can utilize similar solutions to manage their clients' complaints and grievances, ensuring that their company's quality continues to improve.

There is a lack of communication between the government and its citizens. In some parts of the world, the manual process of filing a grievance can take up to a month. Furthermore, tracking the status of a complaint is a time-consuming and tiresome task. With these problems in hand, Jattan *et al.* [64] have used the ethereum blockchain to propose a secure and transparent system for resolving complaints. Each complaint is a smart contract processed on the ethereum Blockchain. The process begins with a user registering a complaint. The complaint is registered and so kept on the ethereum blockchain by including the appropriate details. To track the status of the complaint, the user will be granted a complaint number.

*Table 4.2 Key findings in complaint and grievances management*

| Reference | Key finding | Advantages | Disadvantages |
|---|---|---|---|
| Jattan *et al.* [64] | Proposed a secured and transparent BC system for complaint redressal system using Ethereum. | • Secure<br>• Transparent and immutable | • Applicability<br>• Scalability<br>• Benchmarking |
| Rahman *et al.* [65] | Presented a BC-based anonymous and transparent platform where complainants can submit anonymous complaints and communicate with authorities for resolving their complaints. | • Secure and reliable<br>• Ensures anonymity of users<br>• Efficient<br>• Ensures temper resistance | • Simulation<br>• Privacy<br>• Efficient<br>• Benchmarking |

Officials are able to view the complaint only if the data is stored on the blockchain and take appropriate action on it. The complainant will be kept informed of all actions taken in response to the complaint.

Quick and effective complaint resolution is a critical civic right that every citizen expects from their government. In a typical physical complaint management system, people must physically visit the organization to file a complaint. These complaints are addressed in a committee-based system, which is a really long and tiresome process. There are various online complaint management tools accessible. Most victims are hesitant to submit a complaint through these online platforms due to the lack of transparency.

In [65], Rahman *et al*. have presented a blockchain-based platform for creating an anonymous, transparent, and decentralized environment for complaint and grievance management. Individuals can use this method to lodge anonymous complaints and communicate with the authorities responsible for resolving their problems. This platform enables officials to deal with complaints more efficiently. It makes use of self-sovereign identity and zero-knowledge proof to ensure that users' identities remain anonymous while lodging complaints, thereby reducing their vulnerability to threats. As a storage and sharing mechanism, this platform makes use of the interplanetary file system, which, like blockchain, is resistant to tampering and provides additional security.

**Summary:** To assist future researchers, the essential concepts of existing works have been outlined in Tables 4.2 and 4.3. In contrast to the work of Jattan *et al*. [64], the work of Rahman *et al*. [65] is reliable, efficient, and ensures user anonymity. Despite the fact that both existing works are built on public blockchain, Jattan *et al*. have used ethereum, whilst Rahman *et al*. have used hyperledger aries and hyperledger indy.

Table 4.3 Different utilized properties in complaint and grievances management (– indicates that the required information is missing/not mentioned)

| Reference | Blockchain type | Implementation language & other platform/framework | Consensus algorithm | Blockchain platform |
|---|---|---|---|---|
| Jattan *et al.* [64] | Public | Next.Js, Web3.js | – | Ethereum |
| Rahman *et al.* [65] | Public | – | – | Hyperledger aries and hyperledger indy |

## 4.6 Health insurance claim management using blockchain technology

In the context of health insurance, a claim is any application filed for benefits from the health insurer organization. The client must file a claim so that funds for his or her care can be reimbursed. The health insurance provider must process the claim request in order to manage claims. This means that the insurer must investigate the claim request's authenticity, rationale, and information. There are various stages involved in the processing and management of claims. This process begins with registering the customer, continues with policy issuance, ensures the customer's authenticity by verifying medical certificates [66], keeps customer data confidential, detects false and anomalous claims, and reduces overall management costs [67]. If everything is in order, the health insurance provider reimburses the client's health care provider.

However, the major challenges in claims management include time management, reliance on human supervision, a lack of visibility, and security. Because most management systems rely on human supervision, the claim management and insurance issuance processes are prone to errors. The data collection and processing procedure is entirely dependent on humans, making it prone to inaccuracy [5]. Also, a major constraint with this entire process is the lack of data visibility and availability and security. Consequently, obtaining health insurance and managing claims becomes a laborious and time-consuming process for both the insurer and the insurance provider [5,67]. Another significant issue in insurance claims is that the insurance process is occasionally hampered by service manipulation and misuse by policyholders and providers seeking a higher payment for an insured incident [5]. Because of these concerns, it has become difficult to ensure the integrity and legitimacy of the claim request information [5,67].

With this concern growing in health insurance claims, there is various research work that has focused on building tempering-free claim architectures. Thenmozhi *et al.* [67] have proposed their architecture based on a major objective: develop a blockchain-based health insurance claim processing system that will assist insurance companies in establishing a secure, tamper-resistant network. The proposed system consists of three modules, including:

1. **Registration:** When a person joins the organization by paying the annual premium, the system calculates the required premium. A hospital joins every year and its stay or drop depends upon the voting system. It can enter its offers into the system. Patients can browse offers and choose treatments.
2. **Treatment:** A person chooses a hospital and an offer from the organization's interface (web). The hospital records the person's treatment in the system. The system stores the record.
3. **Claim:** Payment is made in the hospital following treatment. Due to the fact that the record is still open and not closed, the organization calculates the payable amounts. The patient can access his or her personal information, payment history, and payable balance via the user web portal.

Authors from [5] have highlighted the characteristics of blockchain technology that are used for issuing health insurance contracts and claims, including automation, authentication [68], transparency, immutability and decentralization. A layer model is defined based on these characteristics to abstract the functionality of the schema components, namely peer-to-peer mining, blockchain, smart contract and user. As a result, the studies have noted three use cases: *the issuance of an insurance contract (policy), payment of an insurance premium, and claim management.*

Transparency at all levels is absolutely critical to health insurance and more specifically, the health sector. In [31], Sawalka *et al.* have proposed a blockchain-based claim model to ensure transparency between insurance companies, obviating the need for agents and enabling direct contact between insurance companies and hospitals. EthInsurance, the proposed framework, enables efficient and secure access to medical data by patients, providers and other third parties while safeguarding patient privacy.

EthInsurance consists of three modules i.e. *the patient, the hospital and the insurance company*. The blockchain is intrinsically linked to the patient and it determines who is granted access. Three contracts, namely *consensus, permission, and service*, are in charge of all blockchain activity. Experiment results show that EthInsurance is reliable and safe. Utilization of blockchain technology lowers the cost of decentralization. It validates and authorizes data before it is transmitted over the network, minimizing the possibility of unauthorized use of records. Furthermore, each patient has a unique ethureum address and identifier. The authors have used distinct contracts to convey a sense of modularity, which benefits the framework's data security.

**Summary:** An overview of the key contributions, benefits and numerous utilized components of the most significant works is provided in Tables 4.4 and 4.5. All existing mechanisms are secure, reliable, and reliable. By facilitating direct contracts between entities, Sawlka *et al.* [31] have eliminated the necessity for agents. Mendoza-Tello *et al.* [5] have, on the contrary, increased transaction durability. Authors from [67] have reduced network latency in order to construct a safe, tamper-free network.

*Table 4.4 Key findings in claim management*

| References | Key finding | Advantages | Disadvantages |
|---|---|---|---|
| Sawalka *et al.* [31] | Proposed a BC-based system that stream-lines and simplifies the insurance claim process | • Eliminates the need for agents<br>• Enables direct contract between entities<br>• Secure and efficient access to medical data<br>• Reliable and safe | • Enhancement Benchmarking |
| Thenmozhi *et al.* [67] | Developed a BC-based solution to aid insurance firms in establishing a tempering-free secure network | • Secure and reliable<br>• Network latency | • Applicability<br>• Scalability<br>• Security<br>• Benchmarking |
| Mendoza-Tello *et al.* [5] | A layer-based model is built and the three usage scenarios are presented based on the characteristics provided by BC | • Transaction durability<br>• Trust<br>• Reliable and safe | • Feasibility<br>• Benchmarking |

*Table 4.5 Different utilized properties in claim management (– indicates that the required information is missing/not mentioned)*

| References | Blockchain type | Implementation language & other platform/framework | Consensus algorithm | Blockchain platform |
|---|---|---|---|---|
| Sawalka *et al.* [31] | Public | Ganache platform | – | Ethereum |
| Thenmozhi *et al.* [67] | – | Express | Proof of work | Ethereum, hyperledger fabric |
| Mendoza-Tello *et al.* [5] | – | – | – | – |

## 4.7 Health insurance bill management using blockchain technology

Billing is the procedure by which a third party, typically insurance companies, reimburses a patient for treatment delivered by any healthcare provider. This entire process is referred to as the *billing cycle* or *revenue cycle management*, and it incorporates the management of claims, payment, and billing [69].

Claims and billing are core aspects of the health insurance system and are closely intertwined. The whole process starts with the doctor visiting the patient and assigning diagnosis and procedure codes, which are then used by the insurer in order to define the coverage and medical necessity of the services [70]. The patient files a claim with the insurance company, and the insurance company verifies the correlation between the services provided by the healthcare provider and the percentage they must pay. Depending on the outcome, the insurance pays the amount to the healthcare provider on the patient's behalf [71]. The insurer must verify if there is any relevance between the service provided by the healthcare provider, the necessity of those medical treatments and the amount of finances demanded in the claim. The reason behind this procedure is that there can be fraud cases which can be executed by the healthcare provider for higher benefits or even the patient themselves. Customers can deceive insurers by claiming for services that were never delivered, for unneeded medications, for repeated claims, or even for fraudulent insurance cards. Similarly, healthcare professionals may charge a fee for any service or treatment that was not provided, or for providing medically excessive and unneeded services [72]. As a result, billing management is a crucial part of the insurance system for the seamless operation of the whole process. For a long time, health insurance companies managed their whole billing system using pen and paper. Currently, the majority of companies manage claims and billing through the use of digital technologies.

To the best of our knowledge, billing management in health insurance has not been addressed separately over the years. Instead of that, research communities focused on this aspect during fraud and claim management activities. As a result, additional domains have been searched in which researchers have addressed billing management in order to reduce fraudulent activity and streamline the claim process.

Authors from [73] have proposed a blockchain-based approach that uses Internet of Things (IoT) devices for the metering and billing of customer for the electric network. The proposed mechanism aims to solve both rust and privacy issues. They used raspberry pi to simulate metering and hyperledger fabric, with its decentralized structure, to provide transparent and safer solutions. Experimental results show the proposed system is less vulnerable to cyber attacks as a consequence of using blockchain. In addition, by implementing smart contracts and automating all energy tracking procedures, the rate of wrong manual measurements will be reduced.

Sawalka *et al.* [31] have aimed to make the insurance process more seamless and less time-consuming. They use insurance as a mode of payment at the hospital. The hospital directly requests a claim from the company, and the insurance company responds with a claim record after checking details in the patient's records. This system reduces the patient's workload while also keeping the process transparent. This paper's details have been skipped because they have already been discussed in Section 4.6.

**Summary:** Tables 4.6 and 4.7 provide a summary of the findings of the reviewed papers, which might help researchers gain a rapid understanding of what previous researchers have accomplished. Both the mechanisms from Ahmed *et al.* [73] and Sawalka *et al.* [31] have given secure and efficient data access, hence reducing the exposure to cyber threats. In addition to the comparison, Ahmed *et al*. have employed *hyperledger fabric*, while Sawalka *et al*. have employed *ethereum*. As part of the implementation, Ahmed *et al*. utilized *javascript* for their private blockchain, whereas Sawalka *et al*. leveraged the *ganache platform* for their public blockchain.

*Table 4.6 Key findings in billing management*

| References | Key finding | Advantages | Disadvantages |
|---|---|---|---|
| Ahmet *et al*. [73] | A consumer metering and billing system is presented that uses BC technology and IoT devices to address scalability, trust, and privacy concerns | • Scalable energy efficient<br>• Less vulnerability to cyber<br>• attacks<br>• Privacy | • Security<br>• Privacy |
| Sawalka *et al*. [31] | Proposed a BC-based system that streamlines and simplifies the insurance claim process | • Eliminates the need for agents<br>• Enables direct contract between entities<br>• Secure and efficient access to medical data, reliable and safe | • Enhancement<br>• Benchmarking |

*Table 4.7 Different utilized properties in billing management (– indicates that the required information is missing/not mentioned)*

| References | Blockchain type | Implementation language & other platform/framework | Consensus algorithm | Blockchain platform |
|---|---|---|---|---|
| Ahmet *et al*. [73] | Private | Javascript | Proof of concept | Hyperledger fabric |
| Sawalka *et al*. [31] | Public | Ganache platform | – | Ethereum |

## 4.8 Fraud detection and risk prevention in health insurance using blockchain technology

Fraud is the intentional deception or manipulation of information by an individual or organization in order to obtain a financial or personal advantage [74]. Healthcare is a prime target for adversaries to attack and profit from. Among these attacks, healthcare insurance fraud has been a significant source of burden on the healthcare industry in recent years. It has attracted the attention of the government and health insurance companies due to the significant losses it causes them [75]. Faking information, hiding third-party liability, and falsifying electronic bills are all examples of common health insurance fraud. These types of instances can be incredibly harmful to an authentic customer, both physically and financially [30].

The traditional health-care system is built on trust. The patient who provides the insurance card is trusted by the health insurance provider. The service providers, understandably, believe that the patient did not falsify the insurance card. In the event of doubt, they have to go to the insurance company physically. Furthermore, employees cannot be trusted because fraud can occur from within the insurance organization. As a result, the old system cannot be trusted as a system, which leads us back to the issue of trust [75]. The issues of privacy and security are inextricably linked to this trust issue. People are always concerned about their security and privacy before they are concerned about anything else. They will never trust any organization if their security is not ensured. As a result, people's trust in the healthcare system is deteriorating. Also, it is very frustrating considering the time, effort and money wasted on the paperwork of the whole manual process [75]. Calculations indicate that American insurance firms waste up to $375 billion per year as a result of the paperwork and administrative box-ticking [76].

Another significant issue with the current health insurance sector is that the claims are submitted manually to the insurance providers. As a result, if there are potential fraud cases, they often go unnoticed. Furthermore, the manual process often includes private individuals whose responsibilities include detecting fraud and abusive instances and proving them to the government or other regulatory authority. If they achieve their goal, they are financially rewarded, which motivates them to perform their duties more sincerely [77]. Still, the manual system is prone to errors. Many fraud instances go undetected owing to a lack of appropriate proof, resulting in massive economic loss [74]. Additionally, the current manpower and resources available for healthcare insurance are insufficient to prevent these fraudulent instances [30]. Furthermore, policyholders and medical service providers sometimes take advantage of a customer's health insurance benefits through falsification and service abuse to obtain additional funds [19]. In Figure 4.2, different aspects of fraudulent incidents in healthcare insurance have been illustrated.

Taking various forms of healthcare insurance fraud in mind, Liu *et al.* [30] have proposed a blockchain-based anti-fraud system for healthcare insurance that consists of *a cloud platform, a network layer, a core layer, an interface layer, and an application layer*. The main objective of this system is to verify the patient's medical reimbursement request and ensure it complies with the policy's provisions. The proposed architecture includes three blockchain services to address three types of fraud, namely *medical process information inspection*, *third-party responsibility inspection*, and *healthcare insurance bill inspection*. From the experiment results, the authors have shown that the proposed system is tailored to business needs and is based on resolving current real-world problems. It does not require any modifications to the existing information system. Additionally, the architecture is used to call services between systems, decouple systems, and internal modules are also a loosely coupled architecture of building blocks that can adapt to changing business requirements. The capability of solving anti-fraud issues, such as fraudulent data, concealing third-party liability accident fraud, false electronic bill reimbursement and other issues has made the architecture suitable for the current trend of medical information.

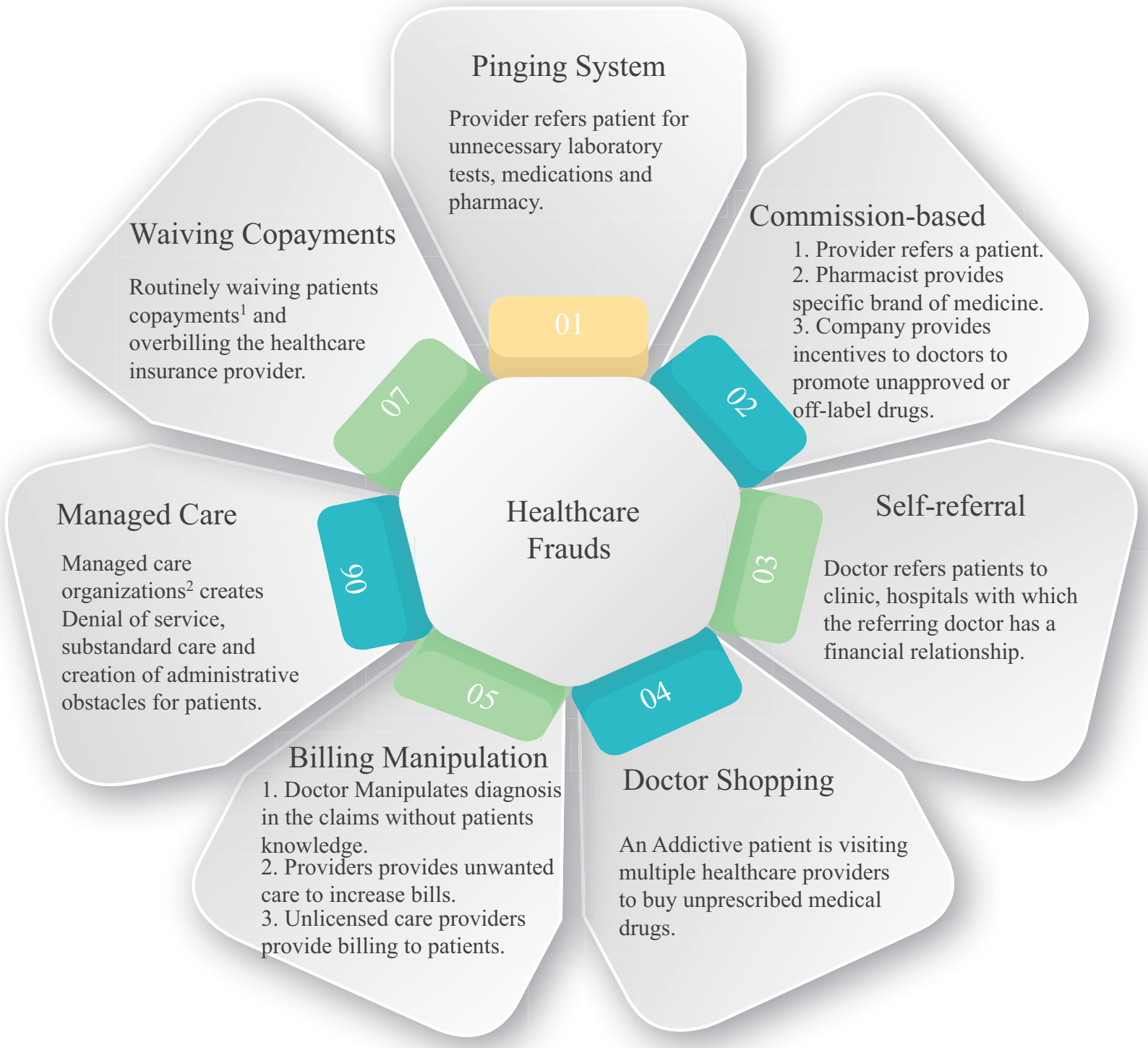

*Figure 4.2 Different aspects of health insurance fraud (inspired by [74]). [1]Patients pay fixed amount to providers as defined by healthcare insurance policy. [2]An entity that connects the healthcare insurance providers and the insurers (patients).*

In [75], Alhasan *et al.* have highlighted the trust between patients and service providers, which leads to counterfeit (fraud) in the health insurance industry. Governments are spending countless amounts of money and time to stop this dilemma. To prevent this, authors have discussed recent articles that looked into various health insurance systems based on blockchain technology and hence proposed a new framework and consensus algorithm to ensure the security and decentralization of a distributed ledger. The proposed BC-based framework is composed of five major components: the insurance company, the patient, treatment providers (hospitals, medical centers, and pharmacies), the ministry of health (MOH), and the blockchain network.

The proposed system is evaluated in terms of validation time, upload time, time required to append blocks to the chain, data integrity verification, and data privacy. Experiment results demonstrate the system's robustness and effectiveness in terms of performance, security, and privacy. The advantages are highlighted below.

- The use of blockchain in storing health information can be effectively secured by having data over multiple machines which are supervised and authorized by a distributed community in preference to a centralized approach.
- This method provides a way for everyone in the party to view and verify the data that is added and modified.
- Moreover, there is a record of each transaction and modification made within the network.
- The performance of middleware to parse and transform medical health data is fast and there is not much observable delay to loading that processed data into blockchain.

Additionally, the proposed consensus algorithm reinforces the distributed concept by selecting randomly based on two rules: FIFO and length results, implying that the system is completely distributed.

Each year, the healthcare industry in the United States loses tens of billions of dollars to fraud. Certain types of fraud put the patient's health at risk. This drives Ismail *et al.* [74] to develop a system capable of detecting and preventing fraud based on twelve different fraud scenarios. Because of the peer-to-peer distributed nature of blockchain, they have proposed *Block-Hi*, a blockchain-based health insurance fraud-detection system. Furthermore, they have created a taxonomy of health insurance fraud based on various fraud scenarios as well as relationships between insurance claim contents, associated fraud categories, and corresponding validators.

The authors have investigated the performance of their system in terms of execution time and data transfer amount when the number of claims and the number of healthcare insurance company branches increased. When the number of health insurance branches increases, the performance of *Block-HI* in terms of execution time and data transferred degrades by only 0.69% on average. While the number of claims have increased, performance declined by 33.51%. This is due to the consensus protocol's execution.

**Summary:** To conclude this section, Tables 4.8 and 4.9 highlight the findings on fraud management using blockchain technology. In addition to providing a blockchain-based solution, Ismail *et al.* [74] have presented a potential taxonomy of fraud attacks in the healthcare insurance industry. The authors have included PBFT into their private blockchain. On the other hand, Liu *et al.* [30] facilitate enhanced system performance. The architecture of Alhasan *et al.* [75] has improved transaction speed and usability. They have introduced a new consensus method for their consortium blockchain.

*Table 4.8 Key findings in fraud detection and risk prevention*

| References | Key finding | Advantages | Disadvantages |
|---|---|---|---|
| Liu *et al.* [30] | Proposed a healthcare insurance anti-fraud system based on BC and cloud computing that reduces the need of resources and manpower and provides various medical services | • Secured<br>• Protects privacy of three-chain data<br>• Supports better system performance | • Enhancement<br>• Benchmarking |
| Alhasan *et al.* [75] | Proposed and implemented a novel framework and consensus algorithm to eliminate fraud in the health insurance industry | • Usability and efficiency<br>• Effective in terms of security<br>• Privacy and speed | • Applicability<br>• Scalability<br>• Benchmarking |
| Ismail *et al.* [74] | Presented a taxonomy of healthcare insurance claim frauds and proposed and evaluated a BC-based healthcare insurance claims fraud detection framework | • Performance enhancement in terms of execution time and the amount of data transferred | • Simulation<br>• Feasibility |

*Table 4.9 Different utilized properties in fraud detection and risk prevention (– indicates that the required information is missing/not mentioned)*

| References | Blockchain type | Implementation language & other platform/framework | Consensus algorithm | Blockchain platform |
|---|---|---|---|---|
| Liu *et al.* [30] | Private | – | – | Hyperledger fabric |
| Alhasan *et al.* [75] | Consortium | C# | Self proposed | – |
| Ismail *et al.* [74] | Private | – | PBFT | – |

## 4.9 Health insurance data record and sharing management using blockchain technology

Patient health data is one of the most powerful weapons in the health insurance sector. As a result of technology breakthroughs such as the invention of the IoT and wearable devices, healthcare data is exploding [78]. This data can provide useful insights to the health insurance company, which can subsequently use it for a variety of purposes, including policy customization and claim management [79,80]. Health insurance companies can use IoT devices to collect data in real-time and build precise and customized policies based on individual lifestyles [80,81]. As a result, it might be said that a secure data sharing infrastructure is required for sharing health data with a healthcare insurance provider in order for the consumer to obtain the full benefits of health insurance.

However, there are several obstacles, such as data privacy, security, and interoperability, that can have a significant impact on the entire data sharing architecture [81,82]. To begin with, there is concern about client data privacy and security, as health data is extremely sensitive by nature and contains personally identifiable information [83]. If this data is compromised, serious financial and physical consequences may occur. As a result, the current data sharing architectures pose a reliability concern [81,82]. Second, effective data integration and interoperability amongst healthcare systems remain a major challenge [81,82]. Another challenge is that customers have little control over their private health data [84]. All of these obstacles necessitate the development of a secure and private infrastructure for data sharing so that data can be shared with proper access control and privacy.

To address the inefficiency and time-consuming nature of the current system, Lokhande *et al*. [79] have devised a new method of sharing health data that makes use of permissioned blockchain technology to safeguard the data and conceal it from those who do not wish to see it. Additionally, authors have utilized the participation service, which is provided by blockchain to assist with distinctiveness management. Their designed mobile healthcare system can collect, distribute and collaborate on individual health data between entities and insurance companies.

The authors of [80] have proposed a method for sharing personal data while maintaining privacy and security through the use of blockchain technology. The proposed novel framework integrates health insurers, IoT-based networks, and blockchain technology to implement an access control protocol based on a smart contract for sharing insureds' financial premiums with stakeholders acting as non-participants or authorized parties. In comparison to traditional data-sharing systems, the proposal evaluation results in authorized access in less time. The proposed method, according to the security analysis, ensures transaction integrity via SHA-256, user authenticity via asymmetric encryption, non-repudiation, avoids single point failure, and privacy.

As the volume of healthcare data increases, issues with unstandardized data formats and provenance become more vulnerable. Lee *et al*. [82] have proposed a standards-based sharing framework, SHAREChain, which incorporates two features to deal with reliability and interoperability issues. These are:

1. It enhances reliability by leveraging the data integrity of a blockchain registry and establishing a consortium blockchain network to exchange data between authenticated institutions.
2. The second feature enhances interoperability with standards relating to healthcare data sharing. To ensure data interoperability, the authors have used FHIR as well as XDS's actor and transaction concepts in the system architecture.

**Summary:** Tables 4.10 and 4.11 provide a comprehensive summary of the overall elaboration of this aspect. Compared to prior publications, [79,80] authors have provided trustworthy and reliable techniques. The mechanism from Iman *et al.* [80] only supports off-chain databases, but the mechanism from Lee *et al.* [82] supports both on-chain and off-chain databases. Iman *et al.*'s technique requires less time due to their usage of PBFT as their private blockchain's consensus algorithm. Lokhande *et al.* have worked with private blockchains as well, but their consensus techniques are top score, proof of veracity, and authentication. SHAREChain, the consortium blockchain technology proposed by Lee *et al.*, is implemented using *XDS* and *FHIR*.

*Table 4.10 Key findings in data record and sharing management (– indicates that the required information is not found)*

| References | Key finding | Advantages | Disadvantages |
|---|---|---|---|
| Lokhande *et al.* [79] | Proposed a system which is capable of exchanging health data by using permissioned blockchain to ensure security and concealment | • Highly ascendable<br>• Trusted and liable | • Benchmarking |
| Iman *et al.* [80] | Presented a novel framework that uses a smart contract to implement an access control system for sharing the financial premiums of insureds with stakeholder as non-participants/ authorized parties | • Protect data from potential threats<br>• Authorized access within less time<br>• Off-chain db | – |
| Lee *et al.* [82] | Proposed a sharing framework which deals with reliability and interoperability issues by utilizing the data integrity and consortium BC network. | • Interoperability<br>• Identity<br>• On/Off-chain | • Emphasize on key factors |

*Table 4.11 Different utilized properties in data record and sharing management (– indicates that the required information is missing/not mentioned)*

| References | Blockchain type | Implementation language & other platform/ framework | Consensus algorithm | Blockchain platform |
|---|---|---|---|---|
| Lokhande *et al.* [79] | Private | – | Top score, proof of veracity and authentication | Hyperledger fabric |
| Iman *et al.* [80] | Private | HTML, CSS, JS, ASP.net, C# | PBFT | – |
| Lee *et al.* [82] | Consortium | XDS, FHIR | – | – |

## 4.10 Limitations and future directions

Blockchain technology has the potential to revolutionize the health insurance sector by resolving complex issues such as claim management and fraud detection. Researchers are recognizing the benefits of integrating blockchain in resolving these challenges of the health insurance sector. Numerous notable research projects have already been suggested and have been demonstrated to be successful in fixing the challenges for which they were established. However, many systems have one or more drawbacks that must be addressed, and additional research must be conducted to solve these limitations.

### *4.10.1 Limitations*

In this subsection, an attempt has been made to highlight the key drawbacks of the reviewed literature in this area. A demonstration of the limitations of the existing literature is given in Table 4.12.

- **Lack of literature:** Blockchain technology is still in its nascent stage. The adoption of this technology is low in comparison to the number of problems it can solve. Despite the fact that the health insurance industry has integrated blockchain in numerous areas, there are still significant areas for which no blockchain-specific solutions exist. Health insurance billing management and complaint and grievance management are two sectors where high-quality blockchain-based solutions are needed. While billing is slightly related to the concept of claiming, it has a significant impact on its own. Although several writers have highlighted the billing process in connection with the claim management process, relatively little study has been conducted on the blockchain-based health insurance billing system. In addition, there is no research on the subject of monitoring and response management, which is highly alarming and should be addressed immediately.
- **Lack of applicability and scalability:** As previously noted, researchers can employ blockchain-based solutions that have already been implemented in other industries to address health insurance challenges. Nonetheless, this poses two crucial questions:

- Are these approaches sufficient for addressing health insurance concerns?
- Are these methods capable of handling health data, which contains sensitive and personally identifying information?

For instance, the authors of [64] have developed a method to control complaints and grievances between the government and the people. They have incorporated blockchain, and the use of blockchain has increased the complaint management system's security, transparency, and immutability. However, they have not provided evidence of the scalability or adaptability of their system, making it extremely difficult to implement the recommended strategy in the health insurance industry or any other sector. In addition, the authors from [67] have failed to clarify the adaptability and performance comparison of their systems, which are critical needs for any insurance firm seeking to use their proposed architecture in a real-world scenario. Alhasan *et al.* [75] have proposed an excellent method for resolving health insurance fraud events, but they have not described the scalability of their method.

*Table 4.12 Limitations founded in reviewed literature (● means limitation exists and ○ means it does not)*

| Limitations in references | Applicability | Scalability | Security | Simulation | Privacy | Feasibility | Enhancement | Emphasize on key factors | Benchmarking |
|---|---|---|---|---|---|---|---|---|---|
| Sawalka *et al*. [31] | ○ | ○ | ○ | ○ | ○ | ○ | ● | ○ | ● |
| Thenmozhi *et al*. [67] | ● | ● | ● | ○ | ○ | ○ | ○ | ○ | ● |
| Mendoza-Tello [5] | ○ | ○ | ○ | ○ | ○ | ● | ○ | ○ | ● |
| Mangaonkar [85] | ○ | ○ | ○ | ○ | ○ | ○ | ○ | ○ | ● |
| Goyal *et al*. [86] | ○ | ○ | ○ | ○ | ○ | ○ | ○ | ○ | ● |
| Jattan *et al*. [64] | ● | ● | ○ | ○ | ○ | ○ | ○ | ○ | ● |
| Rahman *et al*. [65] | ○ | ○ | ○ | ● | ● | ○ | ○ | ○ | ● |
| Ahmet *et al*. [73] | ○ | ○ | ● | ○ | ● | ○ | ○ | ○ | ○ |
| Liu *et al*. [30] | ○ | ○ | ○ | ○ | ○ | ○ | ● | ○ | ● |
| Alhasan *et al*. [75] | ● | ● | ○ | ○ | ○ | ○ | ○ | ○ | ● |
| Ismail *et al*. [74] | ○ | ○ | ○ | ○ | ○ | ● | ○ | ○ | ○ |
| Lokhande *et al*. [79] | ○ | ○ | ○ | ○ | ○ | ○ | ○ | ○ | ● |
| Iman *et al*. [80] | ○ | ○ | ○ | ○ | ○ | ○ | ○ | ○ | ○ |
| Lee *et al*. [82] | ○ | ○ | ○ | ○ | ○ | ○ | ○ | ● | ○ |

Before even considering integrating an existing solution from another industry into the health insurance system, the above two concerns must be answered. Consequently, there is insufficient study to fully comprehend the applicability and scalability of existing solutions in the health insurance industry.

- **Lack of simulation and benchmarking:** Simulation and benchmarking play a crucial part in the success of any scientific study. Most research papers that use blockchain technology to solve problems with health insurance do not have quality benchmarking.

  The authors of [65] offer a significant method for managing complaints and grievances using blockchain technology. However, the authors have not provided a simulation of their proposed system, which is necessary if this strategy is to be successfully integrated into the health insurance system. Similarly, Alhasan *et al.* [75] have offered a practical and effective strategy for addressing the fraudulent concerns that occur in the health insurance market, but they have not presented any benchmarks to evaluate their work. The same restrictions apply to Ismail *et al.* [74], who have not compared the energy consumption and performance of different consensus procedures in Block-HI.
- **Lack of feasibility:** Researchers have presented a variety of potential solutions for health insurance-related problems. However, several of them are missing the feasibility study component. The purpose of a feasibility study is to determine whether the proposed solution to a problem is sufficient. For instance, the approach presented in [5] can be quite effective in the processing of health insurance claims. Though the benefits of incorporating blockchain into claim management are evident, rules cannot be established for every component of any system, especially for the extremely uncertain fraudulent scenarios in claim administration. In some instances, a sophisticated medical treatment must be rigorously validated, which demands a significant deal of experience and expertise. Similarly, Ismail *et al.* [74] have presented a solution to the problem of health insurance fraud. To achieve this, they have divided fraudulent actions into 12 distinct groups. However, manually identifying activities is quite difficult due to the possibility of unanticipated challenges.
- **Security:** System security is a significant consideration while attempting to resolve health insurance problems. A secure environment is one of the grounds for implementing blockchain in health insurance. However, while many of the ideas presented by researchers have the ability to tackle the problems, they also pose significant security risks. The solution given by [67] for the claim management aspect creates serious security problems. The problem with this study is that the database used in the suggested architecture has not been encrypted. The most promising billing management study discovered is a blockchain-based billing system [73] whose concept can be applied into the health insurance billing situation. However, in order to achieve the best results, it is necessary to address the proposed solution's limits first. The authors of this study have deployed rest servers and demonstrated that these can actually accelerate software development; nevertheless, they have not considered server security in order to incorporate higher security for identity management.

In addition to security, user privacy is a particularly delicate issue in any framework that must be addressed prior to using this system for health insurance, as users are highly worried about their health information. The healthcare and health insurance sectors heavily rely on data. However, the nature of health data is so sensitive since it contains personally identifiable information that every individual is exceedingly protective of his or her data. If the system for data sharing is insufficiently secure, privacy breaches can occur, causing serious harm to any individual. Therefore, data privacy and system security are key considerations when offering a solution to health insurance problems.

The authors [65] have presented a way for addressing complaints and concerns in the health insurance business. However, they have neglected to account for user privacy, a crucial component of any complaint and grievance management system. On any online forum for complaint and grievance management, if the user's identity is not anonymous, the majority of users do not feel comfortable sharing their issues. Similarly, although the authors of [73] have offered a solid strategy that emphasizes the billing notion of health insurance, they have not disclosed how they protect their customers' privacy.

- **Lack of accuracy and paperwork:** The success rate of a mechanism is proportional to its accuracy. Therefore, any proposed method must strive for the best level of accuracy possible. Based on their findings, the anti-fraud method for healthcare insurance presented in [30] is quite reliable. However, the accuracy falls short of expectations. Similarly, EhtInsurance [31] is a potential method for the handling of health insurance claims that has been shown to be reliable and secure. However, it lacks sufficient documentation.
- **Lack of emphasis on key factors:** Prior to implementing a solution, the first step in attempting to handle a problem in the real world is to identify the major components that must be addressed. Without it, obtaining maximum achievement is somewhat unpredictable and difficult. Consequently, the first stage in fixing a problem is to identify the crucial parts that must be addressed prior to implementing the solution. However, several of the research have not adhered to this fundamental guideline, which is a major flaw of their system. ShareChain [82] presents a data sharing structure that is mainly absent of this characteristic.

Numerous studies have emphasized the need of secure and effective data sharing in the health insurance industry. ShareChain [82] is one of them, and it has introduced an efficient data-sharing solution to address interoperability and reliability issues. Existing works have been compared to the proposed framework to evaluate their success and limits. The primary issue with this strategy is that it lacks a patient-centered approach. Due to the fact that the customer or patient is at the heart of the healthcare or health insurance industry and is the ultimate owner of their data, the framework should concentrate largely on them. If data collection fails, it is useless to adopt any data-sharing system, and the insurance industry suffers huge losses.

### *4.10.2 Future directions*

In the previous sub section, the limitations of existing blockchain-based health insurance management systems have been highlighted. Based on these limitations, potential future directions have been provided in this subsection.

- **Need for effective solutions:** There is a large gap in the use of blockchain technology to address several areas of health insurance. As indicated in the preceding section, monitoring and response management are lacking in the literature and are not being investigated by researchers. The authors of [87] have provided a monitoring framework that future researchers might use to generate ideas for health insurance monitoring and response management. Perhaps this remains an open question that future scholars should explore. However, there has been a few research on blockchain-based complaint and grievance management systems and blockchain-based billing management systems in numerous areas [88,89]. Using this work, researchers can attempt to integrate these concepts into the health insurance system by resolving the underlying dependencies.

  Compared to the preceding aspect, numerous contributions have been made to claim management and fraud detection in health insurance. Mentioned are several potential approaches that could be utilized in the context of health insurance claim management and the detection of fraudulent activity. The authors of [67] have suggested a blockchain-based strategy for developing a tamper-proof claims processing system for health insurance claims administration. Detecting fraud and preventing risks in health insurance is another area where blockchain has been utilized in multiple research. This chapter looks at the most promising studies from the authors' point of view to give readers a full understanding of how fraud happens and how to stop it.
- **Solutions' applicability and scalability:** A common behavior among health insurance researchers is a lack of awareness or motivation to verify the applicability and scalability of their suggested solutions. There are a few excellent solutions for various aspects of health insurance, such as a complaint redressal system from [64], an anti-fraud system from company [75], and a claim management system from company [67], but none of the architectures have underwent applicability and scalability testing. Future research should be conducted to demonstrate the efficacy and scalability of their proposed method so that it can be applied to the detection of fraud in health insurance and other aspects.

  Regarding the possibility of incorporating an existing solution from another sector into the health insurance system, two considerations are outlined in Section 4.1. Therefore, substantial research is necessary to comprehend the applicability and scalability of existing health insurance systems.

- **Increasing trend in simulation and benchmarking:** As described in the preceding section, simulation and benchmarking play a crucial role in assessing any architecture. If it is unknown whether a newly offered solution will outperform current alternatives, then there is no reason to choose the new one over the old one. This is where simulation and benchmarking come into play, something the majority of studies have lacked. Consequently, future researchers must incorporate simulation and benchmarking prior to deeming their solution beneficial. Additionally, case studies should be presented in exceptional circumstances, such as a consumer emergency complaint registry.
- **Increase feasibility:** For the successful integration of a certain solution, it is essential to ensure that it is feasible. Several of the works that we analyzed lacked this quality. For instance, Ismail *et al.* have developed a very successful method for claim management in health insurance, however it fails to meet some practical situations. As stated previously, authors set up rules for their component, which is not feasible at all. Setting up rules for every component, especially in fraudulent activities, is an unstable state. Critical scenarios such as medical treatment requiring rigorous validation need a lot of expertise and knowledge. Therefore, additional study must be conducted on the integration of machine learning algorithms and blockchain smart contracts. This will aid in the discovery of fraudulent instances and the overall success of the claims management process.

Likewise, Ismail *et al.* [74] have manually categorized fraudulent acts into twelve distinct groups. However, the amount of possible fraudulent actions remains uncertain. In this regard, substantial research is required so that the behavior of various fraudulent operations can be incorporated into future frameworks. Future additions to this proposed system could include an interoperability framework for data claims.

- **Security:** Numerous studies have the potentiality to overcome health insurance management. But their mechanisms raise security concerns which need utmost consideration. For instance, mechanisms from Thenmozhi *et al.* [67] have not made use of encrypted databases. By utilizing fine-grained access control, this concern can be easily accomplished. Ahmet *et al.* [73] have utilized restful servers, which have introduced the topic of finding better servers. This topic needs exhaustive research in order to integrate greater security for identity management. Additionally, post-quantum cryptography can be investigated and included into the process as quantum computers become more widespread as well as a symmetric encryption should be investigated as an alternative.
- **Privacy:** User privacy is a vital aspect in any complaint and grievance management system. In their complaint and grievance management system, Rahman *et al.* [65] have not considered the user's privacy. Therefore, the communication must be made anonymous by utilizing the blockchain's anonymity feature, as this makes the communication more trustworthy, faster, and safe. Besides, Ahmet *et al.* [73] have made no mention of how their proposed billing system would protect their customers' privacy. Thus, planning to integrate a new privacy mechanism is a matter of urgency in these cases because health data is continuous in nature. Privacy techniques such as differential privacy are an excellent strategy for ensuring users' data privacy [83].

- **Increase accuracy and paperwork:** For any proposed method to be effective, a minimum level of precision and essential documentation are required. An anti-fraud system proposed by [30] is a very promising solution to the problem of fraud in the health insurance system, but it lacks a crucial element: accuracy. Subsequently, if additional work is done to improve the system's accuracy, this proposed model could be nearly unbeatable. In addition, written documentation permits us to monitor the development of any project and reveals how to enhance the system. Similarly, EhrInsurance [31] is a promising approach to claim management; however, a major element it lacks is proper amount of paperwork.
- **Emphasize on key factors:** The first and most important stage in proposing a novel solution is to identify the key aspects that must be considered. If this is not ensured, there is a significant chance that the proposed method will become a useless solution. ShareChain [82] is a really promising piece of work that focuses on data sharing in the context of health insurance. However, the most critical issue with this solution is that they have not considered making it patient-centered, which is the most important requirement from a health insurance standpoint. Therefore, it must be taken into account for the system to be functional. The authors have also mentioned that they want to put their system in the real world, which is the ultimate objective of any proposal for a data-sharing platform.

## 4.11 Conclusions

The benefits of blockchain technology in the health insurance industry are evident. With its core qualities of decentralization, persistence, anonymity, and verifiability, blockchain technology has the potential to influence the existing health insurance sector. This chapter discusses numerous aspects of health insurance and how they might interact using blockchain technology. This emerging technology has the ability to minimize the costs associated with a decentralized environment if employed effectively. Numerous studies have examined insurance claims, fraud detection, and data sharing among insurance participants while preserving their privacy and security. Whereas there is a shortage of literature regarding the proper handling of medical bills for insurance claims and the management of complaints within the insurance cycle. Overall, there are very few studies that have explored the various aspects and challenges of health insurance. In addition, future research ought to concentrate on implementing the blockchain-based service in a real world environment in order to assess the practicality and scalability of such a system. Finally, blockchain technology has the ability to alleviate health insurance concerns through decentralization and the elimination of third parties, thereby accelerating the process. Regarding blockchain, health insurers must be fearless. To utilize the unique characteristics of blockchain technology, insurer have to have the courage to develop a new security policy, strengthen the entire process, and contribute to the company's success.

## References

[1] Keisler-Starkey K, Bunch LN. Health Insurance Coverage in the United States: 2020. United States Census Bureau. 2021.

[2] KAGAN J. Health Insurance; Updated March 06, 2022. [Online; accessed April 7, 2022]. https://www.investopedia.com/terms/h/healthinsurance.asp.

[3] What is Health Insurance? [Online; accessed April 7, 2022]. https://www.iciciprulife.com/health-insurance/what-is-health-insurance.html.

[4] Health Insurance – Meaning & Definition; Updated November 04, 2016. [Online; accessed April 7, 2022]. https://www.hdfclife.com/insurance-knowledge-centre/about-life-insurance/health-insurance-meaning-and-types.

[5] Mendoza-Tello JC, Mendoza-Tello T, Villacs-Ramón J. A blockchain-based approach for issuing health insurance contracts and claims. In: *The International Conference on Advances in Emerging Trends and Technologies*. New York, NY: Springer; 2021. p. 250–260.

[6] Levit LA, Balogh E, Nass SJ, *et al. Delivering High-Quality Cancer Care: Charting a New Course for a System in Crisis*. National Academies Press Washington, DC; 2013.

[7] Ellis L, Canchola AJ, Spiegel D, *et al.* Trends in cancer survival by health insurance status in California from 1997 to 2014. *JAMA Oncology*. 2018; 4(3):317–323.

[8] Walker GV, Grant SR, Guadagnolo BA, *et al.* Disparities in stage at diagnosis, treatment, and survival in nonelderly adult patients with cancer according to insurance status. *Journal of Clinical Oncology.* 2014; 32(28):3118.

[9] Ayanian JZ. America's Uninsured Crisis: Consequences for health and health care. Statement before the Committee on Ways and Means, United States House of Representatives Public Hearing on Health Reform in the 21st Century: Expanding Coverage, Improving Quality and Controlling Costs March. 2009. p. 11.

[10] Halpern MT, Ward EM, Pavluck AL, *et al.* Association of insurance status and ethnicity with cancer stage at diagnosis for 12 cancer sites: a retrospective analysis. *The Lancet Oncology.* 2008;9(3):222–231.

[11] Ward E, Halpern M, Schrag N, *et al.* Association of insurance with cancer care utilization and outcomes. *CA: A Cancer Journal for Clinicians.* 2008; 58(1):9–31.

[12] Mandelblatt JS, Yabroff KR, Kerner JF. Equitable access to cancer services: a review of barriers to quality care. *Cancer: Interdisciplinary International Journal of the American Cancer Society.* 1999;86(11):2378–2390.

[13] Zamosky L. Healthcare, Insurance, and You: The Savvy Consumer's Guide. Apress; 2013.
[14] Sommers BD, Gourevitch R, Maylone B, *et al.* Insurance churning rates for low-income adults under health reform: lower than expected but still harmful for many. *Health Affairs.* 2016;35(10):1816–1824.
[15] Health Insurance – Meaning & Definition; Updated November 04, 2016. [Online; accessed April 7, 2022]. https://www.hdfclife.com/insurance-knowledge-centre/about-life-insurance/health-insurance-meaning-and-types.
[16] Chu V. National Health Care Fraud and Opioid Takedown Results; Updated September 30, 2020. [Online; accessed April 7, 2022]. https://www.justice.gov/usao-sdca/pr/national-health-care-fraud-and-opioid-takedown-results-charges-against-345-defendants.
[17] Association NHCAF, *et al. Cost of healthcare frauds.* Accessed: March 2021. p. 11.
[18] Donovan F. Half of US Adults Are Anxious About Healthcare Data Security; July 30, 2018. [Online; accessed April 10, 2022]. https://healthitsecurity.com/news/half-of-us-adults-are-anxious-about-healthcare-data-security.
[19] Kose I, Gokturk M, Kilic K. An interactive machine-learning-based electronic fraud and abuse detection system in healthcare insurance. *Applied Soft Computing.* 2015;36:283–299.
[20] Roy R, George KT. Detecting insurance claims fraud using machine learning techniques. In: *2017 International Conference on Circuit, Power and Computing Technologies (ICCPCT)*. New York, NY: IEEE; 2017. p. 1–6.
[21] Branting LK, Reeder F, Gold J, *et al.* Graph analytics for healthcare fraud risk estimation. In: *2016 IEEE/ACM International Conference on Advances in Social Networks Analysis and Mining (ASONAM)*. New York, NY: IEEE; 2016. p. 845–851.
[22] Srinivasan U, Arunasalam B. Leveraging big data analytics to reduce healthcare costs. *IT Professional.* 2013;15(6):21–28.
[23] Selvakumar V, Satpathi D, Praveen Kumar P, *et al.* Modeling and prediction of third-party claim using a machine learning approach. *Indian Journal of Science and Technology.* 2020;13(21):2071–2079.
[24] IBM. What is blockchain technology?. [Online; accessed April 7, 2022]. https://www.ibm.com/topics/what-is-blockchain.
[25] Namasudra S, Deka GC, Johri P, *et al.* The revolution of blockchain: state-of-the-art and research challenges. *Archives of Computational Methods in Engineering.* 2021;28(3):1497–1515.
[26] HAYES A. Blockchain Explained; Updated March 05, 2022. [Online; accessed April 7, 2022]. https://www.investopedia.com/terms/b/blockchain.asp.
[27] MORRIS DZ. Leaderless, Blockchain-Based Venture Capital Fund Raises $100 Million, And Counting; May 16, 2016. [Online; accessed April 11, 2022]. https://fortune.com/2016/05/15/leaderless-blockchain-vc-fund/.

[28] Economist T. A Venture Fund With Plenty of Virtual Capital, but No Capitalist; October 31, 2015. [Online; accessed April 11, 2022]. https://innovationtoronto.com/2016/05/venture-fund-plenty-virtual-capital-no-capitalist/.

[29] Chowdhury MJM, Usman M, Ferdous MS, *et al.* A cross-layer trust-based consensus protocol for peer-to-peer energy trading using fuzzy logic. *IEEE Internet of Things Journal.* 2021;99:1.

[30] Liu W, Yu Q, Li Z, *et al.* A blockchain-based system for anti-fraud of healthcare insurance. In: *2019 IEEE 5th International Conference on Computer and Communications (ICCC)*. New York, NY: IEEE; 2019. p. 1264–1268.

[31] Sawalka S, Lahiri A, Saveetha D. EthInsurance: a blockchain based alternative approach for Health Insurance Claim. In: *2022 International Conference on Computer Communication and Informatics (ICCCI)*; 2022. p. 1–9.

[32] Contributors E. Overview: Insurance. [Online; accessed April 12, 2022]. https://www.encyclopedia.com/finance/encyclopedias-almanacs-transcripts-and-maps/overview-insurance.

[33] Hamer L. Here's How Health Insurance Companies Ripped Off People For Years; January 15, 2018. [Online; accessed April 12, 2022]. https://www.amazon.com/Bitcoin-Cryptocurrency-Technologies-Comprehensive-Introduction/dp/0691171696.

[34] Cutler DM, Wikler E, Basch P. Reducing administrative costs and improving the health care system. *New England Journal of Medicine*. 2012; 367 (20):1875–1878.

[35] Weinick RM, Burns RM, Mehrotra A. Many emergency department visits could be managed at urgent care centers and retail clinics. *Health Affairs.* 2010;29(9):1630–1636.

[36] Vogeli C, Shields AE, Lee TA, *et al.* Multiple chronic conditions: prevalence, health consequences, and implications for quality, care management, and costs. *Journal of General Internal Medicine.* 2007;22(3):391–395.

[37] Ayanian JZ, Williams RA. Principles for eliminating racial and ethnic disparities in healthcare. In: *Eliminating Healthcare Disparities in America*. New York, NY: Springer; 2007. p. 377–389.

[38] Freeman JD, Kadiyala S, Bell JF, *et al.* The causal effect of health insurance on utilization and outcomes in adults: a systematic review of US studies. *Medical Care.* 2008;p. 1023–1032.

[39] Goodnough A. Census shows a modest rise in US income. *New York Times A*. 2007;1.

[40] Institute of Medicine (US) Committee on the Consequences of Uninsurance I. Care Without Coverage: Too little, Too Late. Washington, DC: National Academy Press; 2002.

[41] Bush G, Carter J, Ford G, *et al. Building a Better Health Care System: Specifications for Reform*. Washington, DC: National Coalition on Health Care; 2004.

[42] Institute of Physicians-American Society of Internal Medicine AC, *et al*. *No Health Insurance? Its Enough to Make You Sick: Scientific Research Linking the Lack of Health Coverage to Poor Health*. Philadelphia, PA: American College of Physicians-American Society of Internal Medicine. 1999.

[43] Brown ME, Bindman AB, Lurie N. Monitoring the consequences of uninsurance: a review of methodologies. *Medical Care Research and Review.* 1998;55(2):177–210.

[44] Sicker HJ. Sicker and poorer—The consequences of being uninsured: a review of the research on the relationship between health insurance, medical care use, health, work, and income. *Medical Care Research and Review.* 2003;60(2 Suppl):3S–75S.

[45] Hoffman C, Paradise J. Health insurance and access to health care in the United States. *Annals of the New York Academy of Sciences.* 2008;1136 (1):149–160.

[46] Howell EM. The impact of the Medicaid expansions for pregnant women: a synthesis of the evidence. *Medical Care Research and Review.* 2001;58 (1):3–30.

[47] Kilbourne AM. Care without coverage: too little, too late. *Journal of the National Medical Association.* 2005;97(11):1578.

[48] Congress U. Office of Technology Assessment: Does Health Insurance Make a Difference. In: *Background Paper*. Washington, DC: US Congress; 1992.

[49] Weissman JS, Epstein AM. The insurance gap: does it make a difference? *Annual Review of Public Health.* 1993;14:243–270.

[50] MarketsandMarkets. Blockchain In Insurance Market by Provider, Application (GRC Management, Death & Claims Management, Identity Management & Fraud Detection, Payments, and Smart Contracts), Organization Size (Large Enterprises and SMEs), and Region – Global Forecast to 2023; July 2018. [Online; accessed April 13, 2022]. https://www.marketsandmarkets.com/Market-Reports/blockchain-in-insurance-market-9714723.html.

[51] Chirag. How Blockchain Technology is Transforming the Insurance Industry; December 16, 2021. [Online; accessed April 13, 2022]. https://appinventiv.com/blog/blockchain-transforming-the-insurance-industry/.

[52] Review AI. How blockchain could revolutionise the insurance industry; Jun 2018. [Online; accessed April 13, 2022]. https://appinventiv.com/blog/blockchain-transforming-the-insurance-industry/.

[53] Amponsah AA, Weyori BA, Adekoya AF. Blockchain in insurance: exploratory analysis of prospects and threats. *International Journal of Advanced Computer Science and Applications.* 2021;12(1):445–466.

[54] Borah MD, Visconti RM, Deka GC. *Blockchain in Digital Healthcare. London:* CRC Press; 2021. Available from: https://books.google.com.bd/books?id=vQdUEAAAQBAJ.

[55] Kuo TT, Kim HE, Ohno-Machado L. Blockchain distributed ledger technologies for biomedical and health care applications. *Journal of the American Medical Informatics Association.* 2017;24(6):1211–1220.

[56] Cheung CF, Lee WB, Wang WM, *et al.* A multi-perspective knowledge-based system for customer service management. *Expert Systems with Applications.* 2003;24(4):457–470. Available from: https://www.sciencedirect.com/science/article/pii/S0957417402001938.

[57] Vickery SK, Jayaram J, Droge C, *et al.* The effects of an integrative supply chain strategy on customer service and financial performance: an analysis of direct versus indirect relationships. *Journal of Operations Management.* 2003;21(5):523–539. Available from: https://www.sciencedirect.com/science/article/pii/S0272696303000627.

[58] Li Z, Guo H, Wang WM, *et al.* A blockchain and AutoML approach for open and automated customer service. *IEEE Transactions on Industrial Informatics.* 2019;15(6):3642–3651.

[59] Wendel S, de Jong JD, Curfs EC. Consumer evaluation of complaint handling in the Dutch health insurance market. *BMC Health Services Research.* 2011;11(1):1–9.

[60] Bendall-Lyon D, Powers TL. The role of complaint management in the service recovery process. *The Joint Commission Journal on Quality Improvement.* 2001;27(5):278–286.

[61] Alkhateeb YM. Blockchain implications in the management of patient complaints in healthcare. *Journal of Information Security.* 2021;12(3):212–223.

[62] Reitsma-van Rooijen M, Brabers A, de Jong J. Bijna 8% wisselt van zorgverzekeraar. Premie is de belangrijkste reden om te wisselen. NIVEL Utrecht; 2011.

[63] Malhotra S, Patnaik I, Roy S, *et al.* Fair play in Indian health insurance. Available at SSRN 3179354. 2018;.

[64] Jattan S, Kumar V, R A, *et al.* Smart complaint redressal system using ethereum blockchain. In: *2020 IEEE International Conference on Distributed Computing, VLSI, Electrical Circuits and Robotics (DISCOVER)*; 2020. p. 224–229.

[65] Rahman M, Azam MM, Chowdhury FS. An anonymity and interaction supported complaint platform based on blockchain technology for national and social welfare. In: *2021 International Conference on Electronics, Communications and Information Technology (ICECIT)*; 2021. p. 1–8.

[66] Namasudra S, Sharma P, Crespo RG, *et al.* Blockchain-based medical certificate generation and verification for IoT-based healthcare systems. *IEEE Consumer Electronics Magazine.* 2023;12(2):83–93.

[67] Thenmozhi M, Dhanalakshmi R, Geetha S, *et al.* Implementing blockchain technologies for health insurance claim processing in hospitals. In: *Materials Today: Proceedings*. 2021. Available from: https://www.sciencedirect.com/science/article/pii/S2214785321019301.

[68] Chowdhury MJM, Chakraborty NR. Captcha based on human cognitive factor. arXiv preprint arXiv:13127444. 2013.
[69] Alex DelVecchio KL. revenue cycle management (RCM); February 2017. [Online; accessed April 13, 2022]. https://www.techtarget.com/searchhealthit/definition/revenue-cycle-management-RCM.
[70] Wikipedia contributors. Medical billing—Wikipedia, The Free Encyclopedia; 2022. [Online; accessed April 9, 2022]. https://en.wikipedia.org/w/index.php?title=Medical_billing&oldid=1071238649.
[71] He X, Alqahtani S, Gamble R. Toward privacy-assured health insurance claims. In: *2018 IEEE International Conference on Internet of Things (iThings) and IEEE Green Computing and Communications (GreenCom) and IEEE Cyber, Physical and Social Computing (CPSCom) and IEEE Smart Data (SmartData)*. New York, NY: IEEE; 2018. p. 1634–1641.
[72] Rivkin JW, Roberto M, Gulati R. Federal Bureau of Investigation, TN - 2001, 2007, and 2009. Harvard Business School Teaching Note 711–487, 2011.
[73] Gür AÖ, Öksüzer S, Karaarslan E. Blockchain based metering and billing system proposal with privacy protection for the electric network. In: *2019 7th International Istanbul Smart Grids and Cities Congress and Fair (ICSG)*. New York, NY: IEEE; 2019. p. 204–208.
[74] Ismail L, Zeadally S. Healthcare insurance frauds: taxonomy and blockchain-based detection framework (block-HI). *IT Professional.* 2021;23 (4):36–43.
[75] Alhasan B, Qatawneh M, Almobaideen W. Blockchain technology for preventing counterfeit in health insurance. In: *2021 International Conference on Information Technology (ICIT)*. New York, NY: IEEE; 2021. p. 935–941.
[76] Mangan D. Health Insurance Paperwork Wastes $375 Billion; January 13, 2015. [Online; accessed April 6, 2022]. https://www.cnbc.com/2015/01/13/health-insurance-paperwork-wastes-375-billion.html.
[77] Bush J, Sandridge L, Treadway C, *et al.* Medicare fraud, waste and abuse. 2017.
[78] Walsh L, Kealy A, Loane J, *et al.* Inferring health metrics from ambient smart home data. In: *2014 IEEE International Conference on Bioinformatics and Biomedicine (BIBM)*. New York, NY: IEEE; 2014. p. 27–32.
[79] Lokhande S, Mukadam S, Chikane M, *et al.* Enhanced data sharing with blockchain in healthcare. In: *ICCCE 2019*. New York, NY: Springer; 2020. p. 277–283.
[80] Hasan IM, Ghani RF. Blockchain for authorized access of health insurance IoT system. *Iraqi Journal of Computers, Communications, Control and Systems Engineering.* 2021;21(3):76–88.
[81] Liang X, Zhao J, Shetty S, *et al.* Integrating blockchain for data sharing and collaboration in mobile healthcare applications. In: *2017 IEEE 28th Annual International Symposium on Personal, Indoor, and Mobile Radio Communications (PIMRC)*. New York, NY: IEEE; 2017. p. 1–5.

[82] Lee AR, Kim MG, Kim IK. SHAREChain: healthcare data sharing framework using Blockchain-registry and FHIR. In: *2019 IEEE International Conference on Bioinformatics and Biomedicine (BIBM)*. New York, NY: IEEE; 2019. p. 1087–1090.

[83] Saifuzzaman M, Ananna TN, Chowdhury MJM, *et al.* A systematic literature review on wearable health data publishing under differential privacy. *International Journal of Information Security.* 2022;21:1–26.

[84] Kish LJ, Topol EJ. Unpatients—why patients should own their medical data. *Nature Biotechnology.* 2015;33(9):921–924.

[85] Mangaonkar OA, Shah D. Health Insurance management process in hospitals using blockchain secured framework. *International Journal of Research in Engineering, Science and Management.* 2021;4(10):77–79. Available from: http://www.journals.resaim.com/ijresm/article/view/1435.

[86] Goyal A, Elhence A, Chamola V, *et al.* A blockchain and machine learning based framework for efficient health insurance management. In: *Proceedings of the 19th ACM Conference on Embedded Networked Sensor Systems. SenSys '21*. New York, NY: Association for Computing Machinery; 2021. p. 511–515. Available from: https://doi.org/10.1145/3485730.3493685.

[87] Agrawal D, Minocha S, Namasudra S, *et al.* A robust drug recall supply chain management system using hyperledger blockchain ecosystem. *Computers in Biology and Medicine.* 2022;140:105100.

[88] Hingorani I, Khara R, Pomendkar D, *et al.* Police complaint management system using blockchain technology. In: *2020 3rd International Conference on Intelligent Sustainable Systems (ICISS)*. New York, NY: IEEE; 2020. p. 1214–1219.

[89] Yaqub R, Ahmad S, Ali H, *et al.* AI and blockchain integrated billing architecture for charging the roaming electric vehicles. *IoT.* 2020;1(2):382–397.